\documentclass[manuscript]{aastex}

\usepackage{pdflscape}

\usepackage{ulem}

\shorttitle{Time dependence of the proton flux}
\shortauthors{Adriani et al.}

\begin{document}
\newcommand{\el}{\mbox{${\rm e^{-}}$}}

\title{Time dependence of the e$^{-}$ flux measured by PAMELA
   during the July 2006 - December 2009 solar minimum.} 

  \author{
O. Adriani$^{1,2}$, G. C. Barbarino$^{3,4}$,
  G. A. Bazilevskaya$^{5}$, R. Bellotti$^{6,7}$, 
M. Boezio$^{8}$, \\
E. A. Bogomolov$^{9}$, M. Bongi$^{1,2}$, V. Bonvicini$^{8}$,
S. Bottai$^{2}$, A. Bruno$^{6,7}$, F. Cafagna$^{7}$, \\
D. Campana$^{4}$,
P. Carlson$^{10}$, 
M. Casolino$^{11}$, G. Castellini$^{12}$, C.~De~Donato$^{13}$, \\
C. De Santis$^{11}$, N. De
Simone$^{13}$, V. Di Felice$^{13,14}$, V. Formato$^{8,\dagger}$, 
 A. M. Galper$^{15}$, \\
A. V. Karelin$^{15}$, 
S. V. Koldashov$^{15}$, S. Koldobskiy$^{15}$, S. Y. Krutkov$^{9}$, \\
A. N. Kvashnin$^{5}$, A. Leonov$^{15}$, 
V. Malakhov$^{15}$, L. Marcelli$^{11}$,  M.~Martucci$^{11,16}$, \\
A. G. Mayorov$^{15}$, W. Menn$^{17}$, 
M Merg\`{e}$^{13,11}$,
V. V. Mikhailov$^{15}$, 
E. Mocchiutti$^{8}$, \\ 
A. Monaco$^{6,7}$,  
N. Mori$^{2}$, R.~Munini$^{8,18}$,
G. Osteria$^{4}$,
F. Palma$^{11,13}$, 
B.~Panico$^{4}$,  \\
P. Papini$^{2}$, M. Pearce$^{10}$, P. Picozza$^{11,13}$, 
M. Ricci$^{16}$,
S. B. Ricciarini$^{12}$, \\
R. Sarkar$^{19,*}$, V.~Scotti$^{3,4}$, M. Simon$^{17}$,
 R. Sparvoli$^{11,13}$, P. Spillantini$^{1,2}$, \\
Y. I. Stozhkov$^{5}$, A. Vacchi$^{8,20}$, 
E. Vannuccini$^{2}$, G. Vasilyev$^{9}$, S. A. Voronov$^{15}$, \\
 Y. T. Yurkin$^{15}$, 
 G. Zampa$^{8}$, N. Zampa$^{8}$,  
M. S. Potgieter$^{21}$, E. E. Vos$^{21}$}
\affil{$^{1}$University of Florence, Department of Physics, I-50019 Sesto Fiorentino, Florence, Italy}
\affil{$^{2}$INFN, Sezione di Florence, I-50019 Sesto Fiorentino, Florence, Italy}
\affil{$^{3}$University of Naples ``Federico II'', Department of Physics, I-80126 Naples, Italy}
\affil{$^{4}$INFN, Sezione di Naples,  I-80126 Naples, Italy}
\affil{$^{5}$Lebedev Physical Institute, RU-119991, Moscow, Russia}
\affil{$^{6}$University of Bari, Department of Physics, I-70126 Bari, Italy}
\affil{$^{7}$INFN, Sezione di Bari, I-70126 Bari, Italy}
\affil{$^{8}$INFN, Sezione di Trieste, I-34149 Trieste, Italy}
\affil{$^{9}$Ioffe Physical Technical Institute,  RU-194021 St. Petersburg, Russia}
\affil{$^{10}$KTH, Department of Physics, and the Oskar Klein Centre for Cosmoparticle Physics, AlbaNova University Centre, SE-10691 Stockholm, Sweden}
\affil{$^{11}$University of Rome ``Tor Vergata'', Department of Physics,  I-00133 Rome, Italy}
\affil{$^{12}$IFAC, I-50019 Sesto Fiorentino, Florence, Italy}
\affil{$^{13}$INFN, Sezione di Rome ``Tor Vergata'', I-00133 Rome, Italy}
\affil{$^{14}$Agenzia Spaziale Italiana (ASI) Science Data Center, Via
del Politecnico snc, I-00133 Rome, Italy}
\affil{$^{15}$National Research Nuclear University MEPhI, RU-115409 Moscow}
\affil{$^{16}$INFN, Laboratori Nazionali di Frascati, Via Enrico Fermi 40, I-00044 Frascati, Italy}
\affil{$^{17}$Universit\"{a}t Siegen, Department of Physics, D-57068 Siegen, Germany}
\affil{$^{18}$University of Trieste, Department of Physics, I-34147 Trieste, Italy}
\affil{$^{19}$Indian Centre for Space Physics, 43 Chalantika, Garia Station Road, Kolkata 700084, West Bengal, India}
\affil{$^{20}$University of Udine, Department of Mathematics and Informatics, I-33100 Udine, Italy}
\affil{$^{21}$Centre for Space Research, North-West University, 2520 Potchefstroom, South Africa}
\affil{$^{\dagger}$Now at INFN, Sezione di Perugia, I-06123 Perugia, Italy}
\affil{$^{*}$Previously at INFN, Sezione di Trieste, I-34149 Trieste, Italy}

\begin{abstract}

Precision measurements of the electron component in the cosmic radiation
provide important information about the origin and propagation of
cosmic rays in the Galaxy not accessible from
the study of the cosmic-ray nuclear components due to their differing
diffusion and energy-loss processes. However, 
when measured near Earth, the effects of propagation and modulation of
galactic 
cosmic rays in the heliosphere, particularly significant for energies
up to at least 30 GeV, must be properly taken into account.
In this paper the electron (\el) 
spectra measured by PAMELA down to 70 MeV 
from July 2006 to December 2009 over 
six-months time intervals are presented. Fluxes are compared with a state-of-the-art three-dimensional model of solar modulation 
that reproduces the observations remarkably well.

\end{abstract}


\section{Introduction}

Electrons are the most abundant negatively charged component of
cosmic-rays but constitute only about 1\% of the total cosmic-ray
flux. Precise measurements of the energy spectrum of cosmic-ray
electrons provide important information for 
the understanding of the origin and
propagation of cosmic rays in the Galaxy that is not accessible from
the study of the cosmic-ray nuclear components. Because of their low
mass, electrons undergo severe energy loss through synchrotron
radiation in the magnetic field and inverse Compton scattering with
the ambient photons. 

There are two prominent origins of high-energy electrons in the cosmic
radiation: primary electrons accelerated at sources such as supernova
remnants, e.g. \citet{all97,aha04}, and
secondary electrons produced by processes such as nuclear interactions
of cosmic rays with the interstellar matter. Additional sources of
electrons such as pulsars, e.g.~\citet{ato95}, or  
dark
matter particles, e.g.~\citet{cir08}, cannot be excluded. Both these
additional sources were invoked to explain the measured 
positron fraction \citep{adr09a,ack12,agu13}. The study of precise
measurements of the energy spectrum of cosmic-ray electrons can shed
light on their origin
and propagation through the galaxy, e.g.~\citet{del10,bis14}. 
However, the majority of the measurements and
the totality of those for energies greater than 100 MeV 
were obtained
with experiments in the proximity of the 
Earth, well inside the heliosphere. Therefore,
the effects of the solar wind and heliospheric magnetic field 
cannot be neglected. 
As cosmic rays traverse the turbulent magnetic field embedded into the
solar wind, particles are scattered by its 
irregularities and undergo convection, diffusion and adiabatic 
deceleration in
the expanding solar wind. Gradient, curvature and current
sheet drifts have also an
effect that is dominant  
during periods of minimum solar activity, 
e.g. see overview \citet{pot13a}. Cosmic rays with rigidities
up to tens of GV are affected but the largest effect is seen at low
rigidities (less than a few GV), e.g. \citet{str14a}. 

In August 2012, Voyager 1 crossed the heliopause, widely considered to
be the modulation boundary and is now inside the very local interstellar
medium \citep{gur13}. For the first time, the very local
interstellar spectra (LIS) at low energies, including the electron LIS
between 5-20 MeV, have been observed, e.g. \citet{sto13,web13,pot14a}.
Together with the PAMELA
measurements at higher energies, these observations make it possible
to properly address a major uncertainty in what the total modulation
of these cosmic rays is between the modulation boundary and the
Earth. 

Furthermore, drift models predict
a clear charge-sign dependence for the modulation of cosmic rays
\citep{pot14b}, whose effects are expected to be particularly evident at
energies below a few GeV.  
During so-called A $<$ 0 polarity cycles like solar cycle $23$, when
the heliospheric magnetic field is directed toward the Sun in the
northern hemisphere, negatively charged particles  
drift inward primarily through the polar regions of the
heliosphere. Conversely, positively charged particles  drift
inward primarily through the equatorial regions 
of the heliosphere, encountering the wavy  heliospheric current sheet
in the process. The situation reverses when the solar magnetic field
changes its polarity at each  
solar maximum, causing in the process a clear 22-year cycle in the
modulation of cosmic rays.

The most recent period of solar minimum activity and the consequent
minimum modulation conditions for cosmic rays were unusual. It was
expected that the new activity  
cycle would begin early in 2008. Instead solar minimum modulation
conditions continued until the end of 2009 when the largest fluxes of
galactic cosmic rays since the 
beginning of the space age were recorded \citep{pot13b,str14b,mew10}. This
period of 
prolonged solar minimum activity is well suited to study the
modulation processes that affect the 
propagation of galactic cosmic rays inside the heliosphere.

Here results on the long-term variation in the energy 
spectrum of galactic cosmic-ray electrons (\el) measured down to 70
MeV are 
presented. These results 
are based on the data set collected by
the PAMELA (Payload for Antimatter Matter Exploration and Light-nuclei
Astrophysics) satellite-borne 
experiment~\citep{pic07} between July 2006
and December 2009.
PAMELA is an instrument designed for cosmic-ray
antimatter studies and is flying on-board the Russian Resurs-DK1 satellite
since June 2006, in a semi-polar  
near-Earth orbit.
Results on the effects of the solar
modulation on the energy spectra 
of galactic cosmic-ray protons in the same period 
have already been published~\citep{adr13a}, with accompanying
numerical modelling by \citet{pot14c}.

\section{The PAMELA instrument \label{sec:instrument}}

\begin{figure}[t]
\centering 
\includegraphics[width=7.cm]{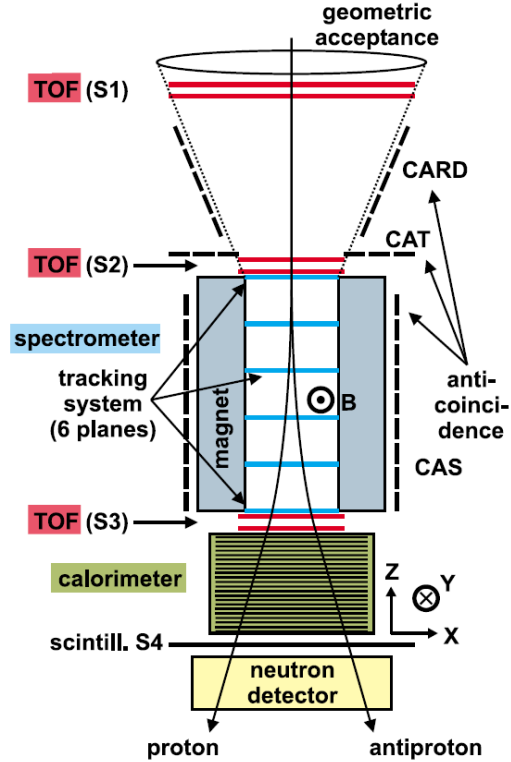}
\caption[]{Schematic view of the PAMELA apparatus. }
\label{fig:appa}
\end{figure}

The PAMELA spectrometer \citep{pic07} was designed and built to study
the antimatter component of cosmic rays from tens of MeV up to
hundreds of GeV and with a significant 
increase in statistics with respect to previous experiments. To achieve
this goal the apparatus was optimized for the study of charge one
particles and to reach a high level of 
electron-proton discrimination.
The instrument, shown schematically in Figure~\ref{fig:appa}, 
comprises the following subdetectors (from top to bottom): a
Time-of-Flight system (ToF S1, S2, S3);  
a magnetic spectrometer; an anticoincidence system (CARD, CAT, CAS);
an electromagnetic imaging calorimeter; a shower tail catcher
scintillator (S4) and a neutron 
detector.
These components
are housed inside a
pressurized container attached
to the Russian Resurs-DK1 satellite, which was launched on
June 15$^{{\rm th}}$ 2006. The orbital altitude varied between
350~km and 600~km at an inclination of 70$^\circ$. 

The central components of PAMELA are a permanent magnet and a tracking
system composed of six planes of double-sided silicon sensors, which
form the magnetic 
spectrometer \citep{adr03}. 
The main task of the magnetic spectrometer is to measure
the particle rigidity R =$ pc/Ze$ ($p$ and $Ze$ being the
particle momentum and charge, respectively, and $c$ 
the speed of light) and the ionization energy losses (dE/dx). The
rigidity measurement is done through the reconstruction of the
trajectory based on the impact points on  
the tracking planes and the resulting determination of the curvature
due to the Lorentz force.  
The ToF system \citep{ost04} comprises three double layers of
plastic scintillator 
paddles  with the first two (S1 and S2) placed above and the third (S3) 
immediately below the magnetic spectrometer, as shown in
Figure~\ref{fig:appa}.  
The ToF system provides the 
measurements of the particle velocity combining the time of passage
information 
with the track length derived from the magnetic spectrometer. By
measuring the particle 
velocity, direction and curvature the spectrometer can distinguish 
between down-going particles and up-going splash-albedo particles 
and separate negatively from positively charged
particles.  

The sampling imaging calorimeter ($16.3$ radiation lengths, $0.6$
interaction lengths) is used for hadron-lepton separation, using
topological and energetic  
information about the shower development in the calorimeter
\citep{boe02}. The shower 
tail catcher and the neutron detector \citep{sto05} beneath provide
additional 
information for the discrimination. An anticoincidence system is used
to reject spurious event \citep{ors05}.

The total weight of PAMELA is 470
kg while the power consumption is 355 W. A more detailed description
of the instruments 
and the data handling can be found in \citet{pic07}.

\section{Data analysis}

This work is  
based on data collected between July~2006 and December~2009. The
periods of time spent by the satellite in the
South Atlantic Anomaly and during significant solar activity
(hence December 2006, when a large solar event
took place \cite{adr11a}) were excluded from the data. 
Data are presented in six-month time periods, a compromise between
statistically significant results and detailed analysis of the time
variation of the fluxes.

\subsection{Electron selection}
\label{s:sel}
Clean events were selected requiring:
\begin{description}
\item[1] A single track fitted within the spectrometer fiducial volume
where the reconstructed track is at least 1.5 mm away from the magnet walls.
\item[2] Selected tracks must have at least 
three hits on the bending $x$-view, at least three hits on the
non-bending 
$y$-view and a track lever-arm of at
least four silicon planes in the tracker. 
\item[3] A positive value for the velocity $\beta = v/c$ ($v$
  particle velocity, $c$ speed of light) measured by
  the ToF system.
\end{description}
This 
set of basic criteria 
provided events with reliable measurements of the 
sign and absolute value of the particle rigidity and velocity. 
\begin{figure}[t]
\centering 
\includegraphics[width=15.cm]{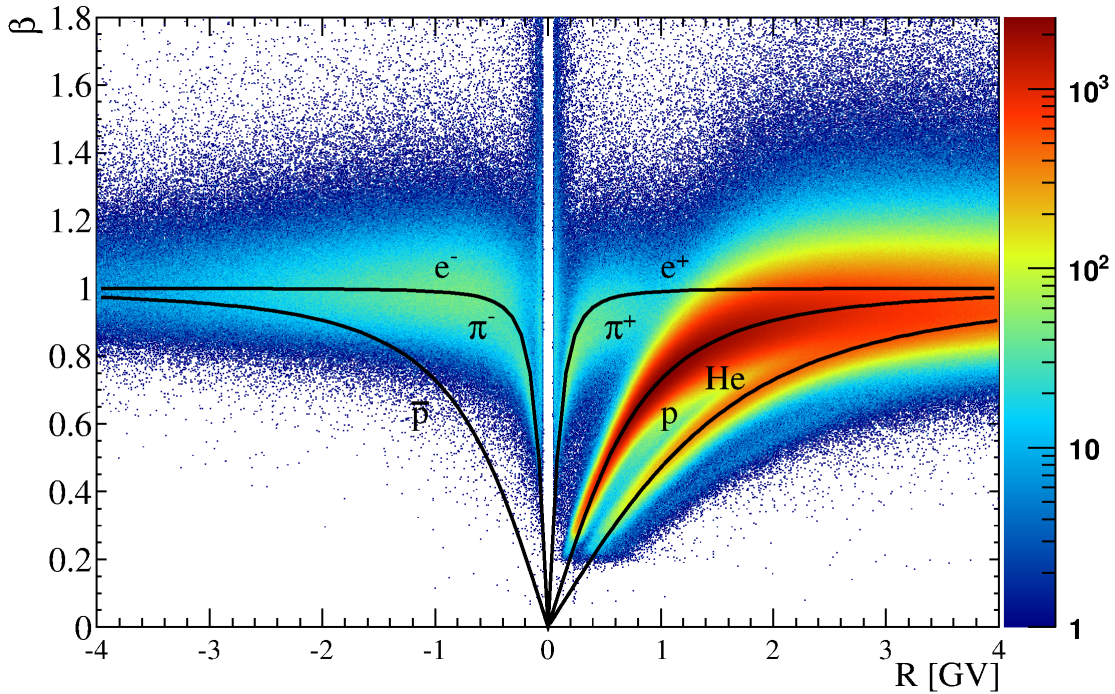}
\caption[]{ToF velocity ($\beta$) as a function of the rigidity. The
  various particle species are indicated. }
\label{fig:beta}
\end{figure}
Figure~\ref{fig:beta} shows the distribution of the velocity \textbf{($\beta$)} as
function of rigidity for 
these events. The spread in the values of $\beta$ for relativistic particles is due to the finite time resolution of the ToF system.
On the positive side the proton signal dominates, while on the
negative side the electron signal clearly emerges as relativistic
particles. However, additional particle species are present in the
negatively charged sample. They are: antiprotons, pions and
``spillover'' protons. The galactic antiproton component represents a
contamination of a few percent over the entire rigidity range.
The pion component is clearly visible below 300~MV in
Figure~\ref{fig:beta} both for positive and negative rigidities. This
component had already been studied
for the antiproton analysis \citep{adr09b} using both simulated and
flight data. The
majority of these pion events had hits in the AC 
scintillators and/or large energy deposits in 
one of the top ToF scintillators clearly
indicating that they were the product of cosmic-ray interactions with
the PAMELA structure or pressure vessel. Spillover protons were mostly
relativistic events
with incorrect determination of the charge sign. These events included\textbf{:}
high energy protons to which the wrong sign of the curvature was
assigned due to the intrinsic deflection uncertainty in spectrometer
measurements\textbf{,} protons that scattered in the 
material of the tracking system mimicking the trajectory of
negatively charged particles and events with spurious hits in the
tracker planes causing a wrong reconstruction of the curvature. The
last two effects were the dominant causes for 
protons reconstructed with low negative rigidities.
This contamination was particularly
significant at very low rigidities (below $\sim 500$~MV) where noisy
strips could be taken as good points for the fit of a highly bent
track when the minimum
requirement on the number of hits on the $x$-view was just three, as
in Criterion 2. For this reason, a 
more stringent criterion
was used in place of Criterion 2 to evaluate the electron fluxes below
500 MV: 
\begin{description}
\item[2bis] Selected tracks must have at least 
four hits on the bending $x$-view, at least three hits on the non-bending
$y$-view and a track lever-arm of at
least four silicon planes in the tracker.  
\end{description}

Then, additional selection criteria were introduced to select as clean as
possible sample of electrons:
\begin{description}
\item[4] No activity in the CARD and CAT scintillators of the anticoincidence system below 10 GV, and no activity in the CAS scintillators below 300 MV.
\item[5] Mean $dE/dx < 3$~mip (minimum ionizing particle units)  in both ToF S1 and S2 scintillators.
\item[6] Mean ionization energy losses (dE/dx) in the tracking system
  planes less than 1.8 mip 
\item[7] Relativistic particles: $\beta > 0.9$.
\item[8] Calorimeter selections.
\end{description}

Criteria 4-5 significantly reduced the pion contamination. The rigidity
ranges for the anticounter selection were a compromise between
residual pion contamination and electron selection efficiencies. As
the electron energy rises, back-scattering from the electromagnetic
shower in the calorimeter increases resulting in an increasing
activity in the anticoincidence scintillators. The different rigidity 
limit for CARD and CAT respect to CAS was due to the different
location of the scintillators respect to the calorimeter (see
Figure~\ref{fig:appa}). 

Criteria 6-7 were used to reduce the antiproton and pion
contaminations 
to a negligible amount up to about 1.7 GV and about 
250 MV, respectively. The residual pion and antiproton contaminations
at higher rigidities and the spillover proton contamination were 
removed using the calorimeter information (Criterion 8).

\begin{figure}[t]
\centering 
\includegraphics[width=14.cm]{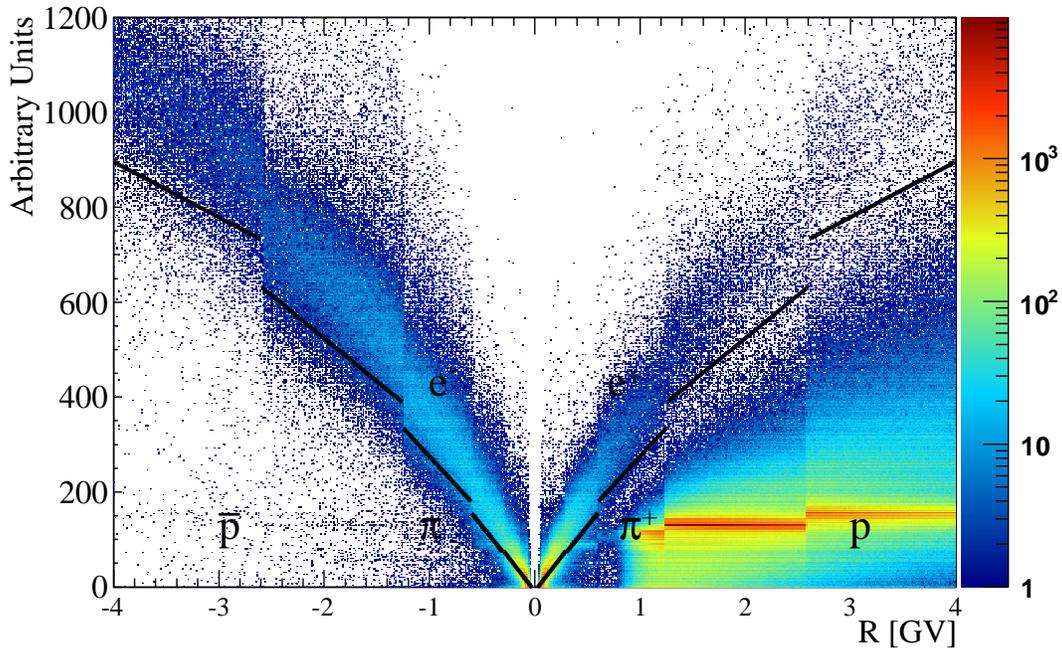}
\caption[]{A quantity related to the topological development of the
  shower in the calorimeter as a function of rigidity for events
  selected with Criteria 1-7. The quantity computation uses the number of the plane closest to the shower maximum estimated for an electromagnetic
  shower of a given energy. The quantization of the plane numbers produce the shown discontinuities.
The events above the solid lines are tagged as electrons by this
selection. }
\label{fig:Calo}
\end{figure}

The calorimeter selection was developed using a Monte Carlo
simulation of the PAMELA apparatus based on the GEANT4 code
\citep{ago03}. The simulation 
reproduces the entire PAMELA apparatus, including the pressure vessel,
and was validated using particle beam data. 
\begin{figure}[t]
\centering 
\includegraphics[width=14.cm]{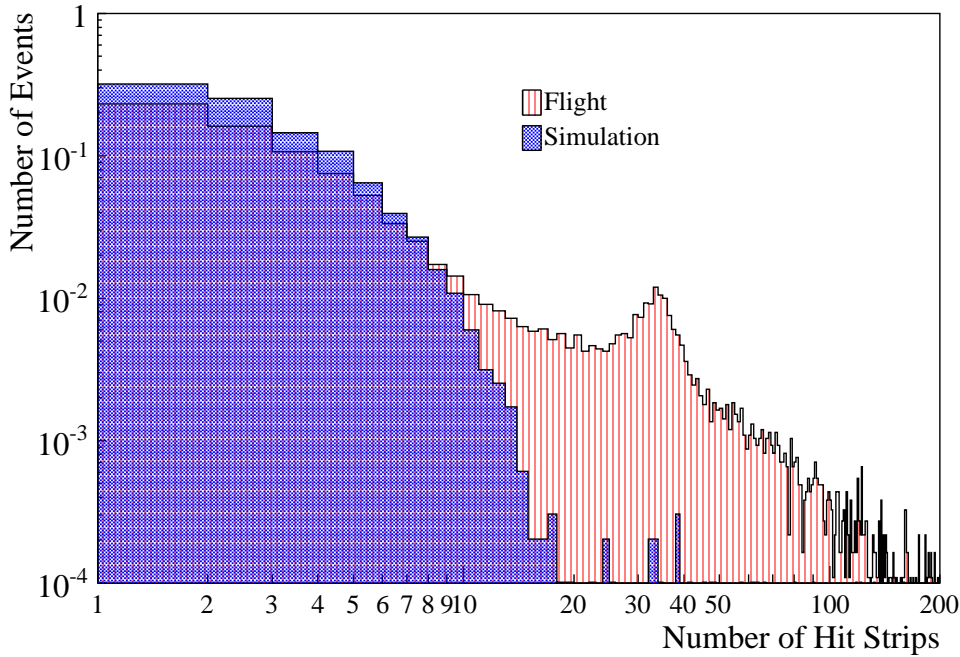}
\caption[]{The normalized number of hit strips in the last 18 calorimeter 
planes for events selected with Criteria 1-7 and rigidity between 70
and 150 MV for experimental (red histogram) and simulated 
(blue histogram) data. The total number of events in both histograms 
is normalized to 1.
The tail to higher values in the number of hit strips  
for the experimental histogram is associated with 
a contamination of spillover protons traversing most of the 
calorimeter, with the
peak around 35 due to non-interacting ones. }
\label{fig:Calo2}
\end{figure}
The longitudinal and transverse segmentation of the calorimeter
allowed 
leptonic showers to be selected with high efficiency and small
contamination above 300 MV. This information was used in previous
analysis to successfully select positrons in a vast background of
protons \citep{adr09a,adr10,adr13b}.  
The calorimeter electron selection was based on variables that
emphasized the differences between the  
leptonic and hadronic shower like the multiplication with increasing
calorimeter depth and the collimation of the electromagnetic cascade
along the track. 
Figure ~\ref{fig:Calo} shows the distribution of one of these variables
for the events surviving Criteria 1-7.
This quantity, related to the multiplication of the leptonic shower,
turned out to have large values for leptons,  
lower for non interacting and late interacting hadrons because of the
limited number of secondaries in the hadronic shower.
The solid lines in Figure ~\ref{fig:Calo} indicate the lower limit for
electron selection based on this quantity. 
Combining several of these variables all residual
contaminations were reduced to a negligible ($\ll 1\%$) amount from 350
MV up to 
the highest rigidities of this analysis (see also \citet{mun12}).   
 
 \begin{figure}[t]
\centering 
\includegraphics[width=14.cm]{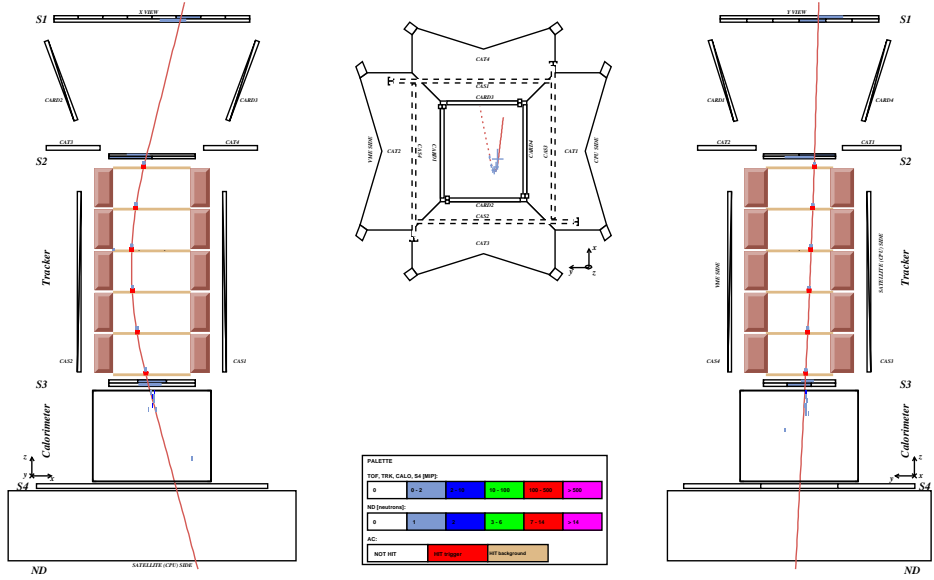}
\caption[]{The event display of an electron event of $\sim$120~MV 
  selected as having 
  a low value in the experimental distribution of Figure~\ref{fig:Calo2} .
The bending (x) and non-bending (y) views are shown on the left and on
the right, respectively. 
A plan view of PAMELA is shown in the center. 
The signal as detected by PAMELA detectors are shown along with the
particle trajectory (red solid line) reconstructed by the fitting
procedure of the tracking system. }
\label{fig:el}
\end{figure}

At the lowest rigidities (below 350 MV) 
spillover protons accounted for most of the residual contamination 
after selection with Criteria 1-7 and were
rejected using additional calorimeter variables that  
exploited the energy deposit in the bottom part of the
calorimeter. Electromagnetic showers below 
about 0.5 GV mostly develop in
the first half of the calorimeter, while spillover protons tend to 
traverse the entire volume. These features are illustrated in
Figure~\ref{fig:Calo2} that shows   
the number of strips hit in the last 18 calorimeter
planes for events selected with Criteria 1-7 and rigidity between 70
and 150 MV
from simulated (blue histogram) and experimental (red histogram) data. 
The tail to higher values in the red histogram is associated with 
a contamination of high energy particles traversing most of the 
calorimeter, with the
peak around 35 due to non-interacting particles, 
while electron-like events account for  
the part of the distribution at low values as indicated by the consistency 
with the Monte Carlo data. This association is further 
confirmed by a visual inspection of events from the 
experimental distribution.
 \begin{figure}[t]
\centering 
\includegraphics[width=14.cm]{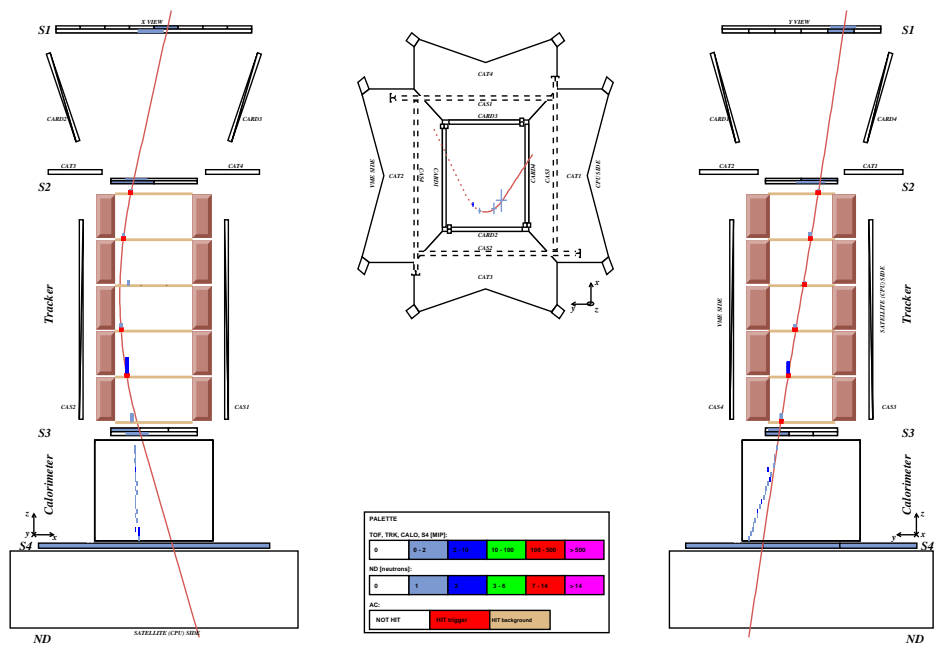}
\caption[]{The event display of a spillover proton event 
of $\sim$110~MV selected from the peak around 35 in 
the experimental distribution of Figure~\ref{fig:Calo2} .
   See Figure~\ref{fig:el} for further details. }
\label{fig:pr}
\end{figure}
Figures~\ref{fig:el} and \ref{fig:pr} show two 
typical events:  
one with a value of 2, consistent with an electron signal, and one with a
value of 32, consistent with a contaminating spillover proton. 
A selection based on this and two related quantities 
rejected this type of contaminating events without affecting
significantly the 
electron signal. 

 \begin{figure}[t]
\centering 
\includegraphics[width=14.cm]{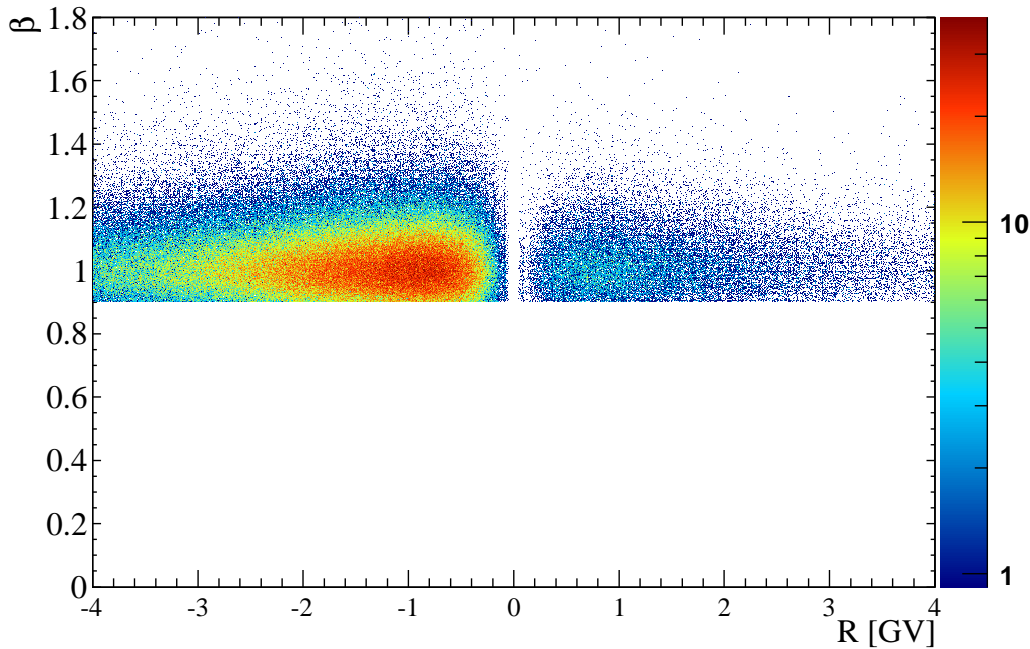}
\caption[]{ToF Velocity ($\beta$) as a function of the rigidity for the
  events surviving all selection criteria, included Criterion 7: $\beta > 0.9$. }
\label{fig:beta2}
\end{figure}

Figure~\ref{fig:beta2} shows the distribution of the velocity as
function of rigidity for the events surviving all selection criteria. 
The residual contamination of pions,
antiprotons and spillover protons was assumed negligible over the
entire rigidity range of interest for this work.  

\subsection{Efficiency}

\begin{figure}[t]
\centering 
\includegraphics[width=14.cm]{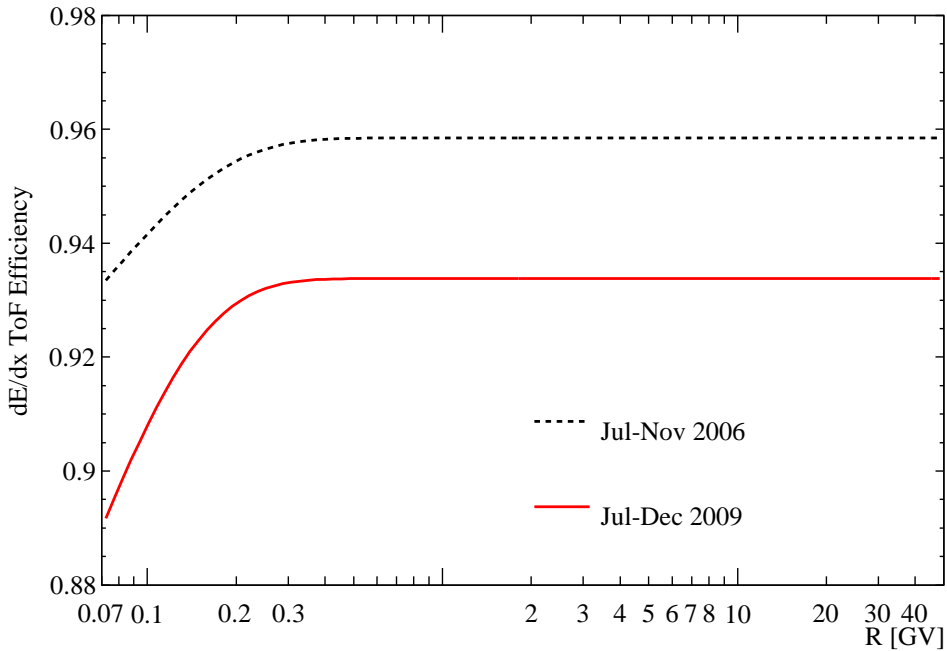}
\caption[]{Temporal evolution of the ToF dE/dx selection
  efficiency. Black dashed 
  line: efficiency for the first time interval (July-November
  2006), red solid line: efficiency for the last time interval
  (July-December 2009). }
\label{fig:Effdedx2}
\end{figure}

As can be seen in Figure~\ref{fig:beta}, the
majority of the negatively charged events were electrons. Along with
the redundant 
information provided by the apparatus, this allowed the study of the
electron selection efficiencies to be conducted 
using flight data. Furthermore, the
large collected 
statistics allowed the time dependence of the efficiencies to be
monitored 
over relatively short time scale. The efficiency study was
complemented by an analysis of simulated data. 
With the Monte Carlo data it was possible to reproduce and study all
selection efficiencies, their 
rigidity and time dependence allowing also the detection of 
possible sources of bias in the experimental evaluation of the
efficiencies, like 
contamination of efficiency samples and correlation among selection
criteria.  
 \begin{figure}[t]
\centering 
\includegraphics[width=14.cm]{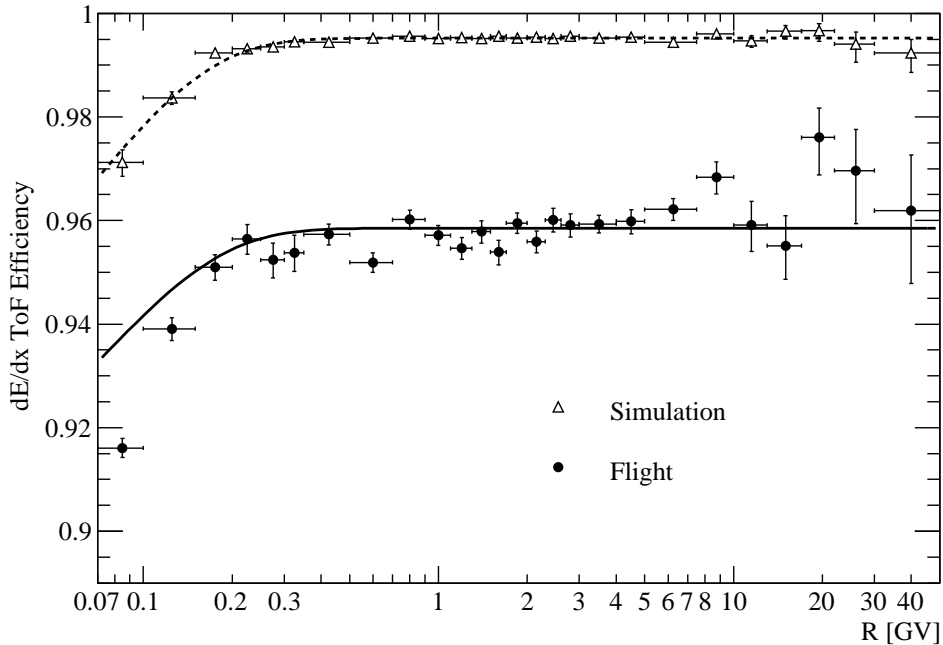}
\caption[]{Efficiency of the ToF dE/dx selection (Criterion 5) as a
function of 
  rigidity for the first time interval (July-November 2006) for flight
  (full circles) and simulated (open triangles) data. The dashed
  line is a fit to the simulated data; the solid line is a fit to the
  experimental data based on the simulated shape and indicates the
  efficiency used in the data analysis. }
\label{fig:Effdedx}
\end{figure}
As an example Figure ~\ref{fig:Effdedx} shows the efficiencies for
the ToF dE/dx selection (Criterion 5) 
for the first time period (July-November
2006) as a function of 
rigidity. The efficiency sample, both experimental and simulated, was
selected using all other selection criteria but Criterion 5. Monte Carlo
data showed that the ToF dE/dx selection efficiency was unaffected by
the selections used to extract the efficiency sample. 
The full circles indicate 
the estimated experimental electron selection efficiency and the open
triangles the simulated one. A slight difference ($\sim 3\%$) can be
seen between the two sets of data, however it should be noted that the
shape of the flight data is well reproduced by simulation except at
very low rigidities, below about 150 MV. In this rigidity region the
difference between experimental and 
simulated efficiencies increases to about 5\% at 70 MV. This additional
difference was due to a residual contamination in the experimental
efficiency sample, as shown by a visual inspection of a random sample
of events. In fact, it was noticed that only a combination of
selections based on 
all PAMELA detectors was able to produce a clean electron sample at
the lowest rigidities. 
Therefore, the ToF dE/dx selection efficiency was obtained fitting the
flight data (solid line in
Figure~\ref{fig:Effdedx}) with a functional shape based on the
simulated data (dashed line in 
Figure~\ref{fig:Effdedx}).  
Figure~\ref{fig:Effdedx2}
shows the resulting efficiency for Criterion 5 at the
beginning and at the end of the data taking.
A small time dependence (about 2\% in nearly four years) of the
efficiency can be noticed.   

Similarly to the case of the analysis of the 
proton flux \citep{adr11b,adr13a}, the efficiency of the tracking system
selection (Criteria 1 and 2) and, especially, its 
energy dependence was obtained 
by Monte Carlo data. 
The tracking system selection efficiency was found to decrease over
the years 
from a maximum of $\sim 90\%$ in 2006 to $\sim 20\%$ at the end of
2009 when Criterion 2 was used in the selection, with Criterion 2bis
the decrease in the efficiency was sharper, down to $\sim 10\%$ at the
end of 2009.
This significant time dependence was due to the sudden, random
failure of a few 
front-end chips in the tracking system. 
This resulted in a progressive reduction of
the tracking efficiency, since the number of hits available for track
reconstruction decreased. However, no degradation in the
signal-to-noise ratio 
and spatial resolution was observed. The front-end chips failure was
treated in the simulation with the inclusion of a
time-dependent map of 
dead channels. 

Another exception was the anticounter selection (Criterion 4)
efficiency for which the simulated values were used. While there was an
excellent agreement between the experimental and simulated
efficiencies, the Monte Carlo predicted a dependence of the efficiency
on the shape of the electron energy
spectrum when measured as a function of rigidity in the spectrometer
instead of energy at 
top of the payload. Considering that the electron 
spectral shape varied significantly over
the orbit due to the Earth's magnetic field (see 
section~\ref{s:cutoff}), it was decided to correct 
with the simulated efficiency  
the selected events distributed according to
their energies reconstructed at the top of the payload, i.e. the
unfolded count distribution, see 
next section.

 \begin{figure}[t]
\centering 
\includegraphics[width=14.cm]{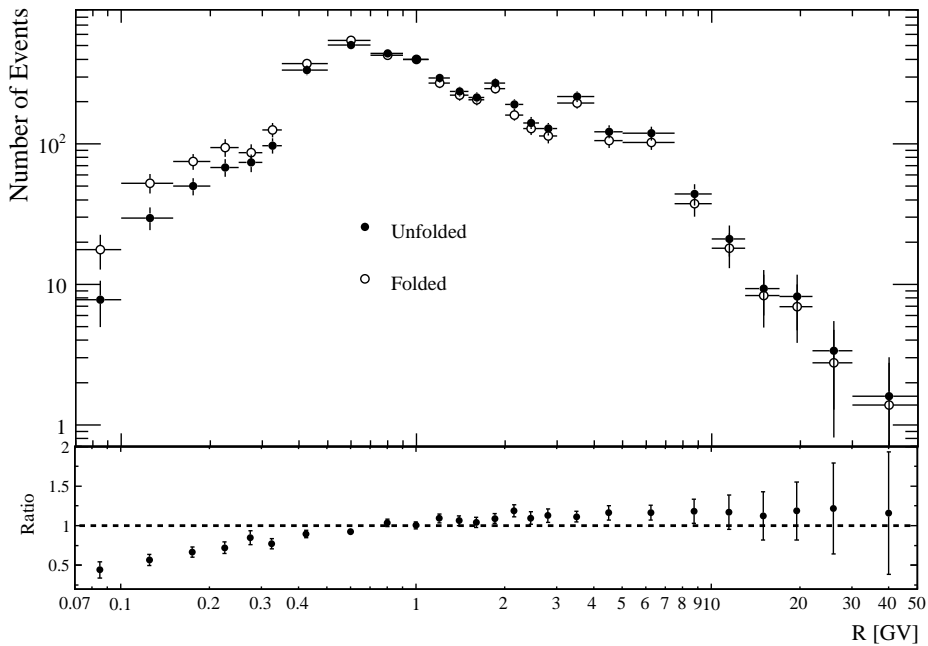}
\caption[]{Top panel: distributions of the event counts,
  corrected for all selection efficiencies except those of selection
  Criteria 2 and 4, selected in the lowest 
geomagnetic cutoff interval ($0-0.055$~GV) before (open circles)
and after (full 
circles) the unfolding procedure. Bottom panel: ratio between the
unfolded and folded count distributions. }
\label{fig:unfold}
\end{figure}

\subsection{Spectral unfolding}

Since in this analysis the electron energies were obtained by measuring
the deflections, 
hence the rigidities, of the particles in the magnet cavity, both the
response of the spectrometer and  
the energy losses suffered by the electrons prior entering the tracking
system had to be properly
accounted for. Particularly significant were energy losses 
due to bremsstrahlung of electrons while
traversing the pressurized container and parts of the apparatus on top
of the tracking system (equivalent to about 0.1
radiation lengths), since the resulting photons were 
able to traverse the spectrometer without being
detected. Consequently the measured rigidities differed from the initial
energies of the electrons at the top of the payload. To account for
these effects a Bayesian 
unfolding procedure, as described in~\citet{dag95}, was applied to the
count distributions of selected 
events binned according to their measured rigidities and divided by
all selection efficiencies except those of the tracking
system and anticounter 
selections. As discussed in the previous section, these were, instead,
applied to the  unfolded count distribution.
Figure~\ref{fig:unfold}, top panel, shows the count distribution for
the lowest 
geomagnetic cutoff interval ($0-0.055$~GV) before (open circles)
and after (full 
circles) the unfolding procedure. The bottom panel shows the variation
of counts in each rigidity (in the spectrometer)/energy (at the top of
the payload) bin resulting from the unfolding
procedure.

\subsection{Flux Determination}
\label{s:cutoff}

The fluxes $\phi(E)$ (E kinetic energy) were evaluated as
follows:

\begin{equation}
  \phi(E) = \frac{N(E)}{\epsilon(E) \times G(E) \times T \times
    \Delta E} 
\end{equation}

where $N(E)$ is the unfolded count distribution, $\epsilon(E)$ the
efficiencies of the remaining tracking system and anticounter
selections, $G(E)$ the 
geometrical factor, $T$ the live-time and $\Delta E$ the width of  
the energy interval. 

The geometrical factor, i.e. the requirement of triggering and
containment, at least 1.5~mm away from the magnet
walls and the TOF-scintillator edges,  
was estimated with the full simulation 
of the apparatus. Hence, it accounted for the geometry of the
instrument, the magnetic field and all physical processes such as
energy losses, multiple scattering, etc.. It was found 
to be
constant at 19.9 cm$^{2}$~sr above 1 GeV, decreasing smoothly to 8 
cm$^{2}$~sr at 70 MeV. This decrease was due to the 
curvature of electrons in the magnetic spectrometer. The PAMELA 
instrumental limit 
for electrons is $\simeq 47$~MeV, below which the
particle trajectory hits the magnet walls.

 \begin{figure}[t]
\centering 
\includegraphics[width=14.cm]{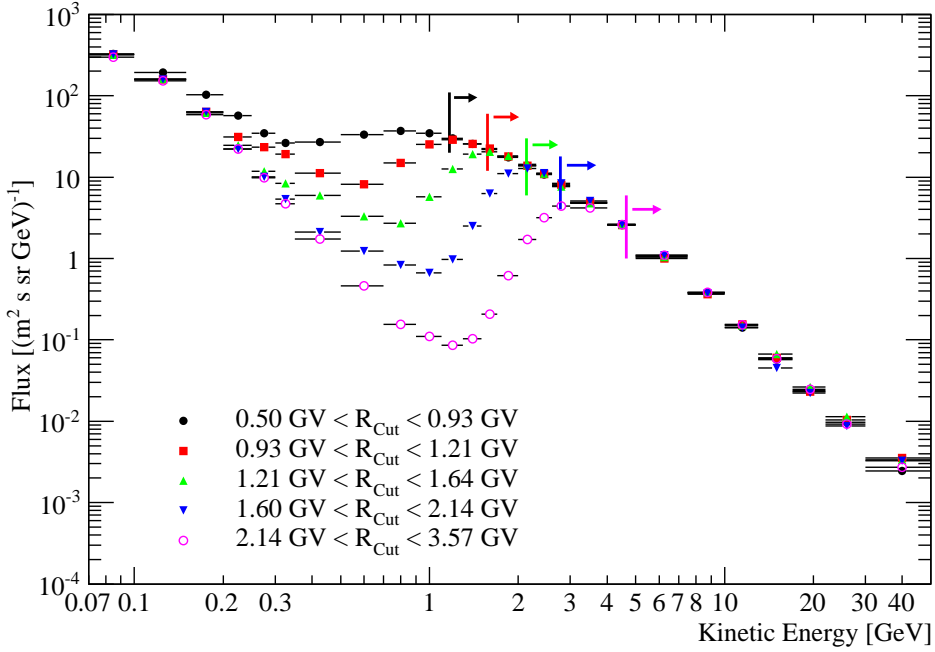}
\caption[]{Electron (\el) energy spectrum measured by PAMELA at the five 
   geomagnetic rigidity cutoff ($\rm{R_{Cut}}$) intervals specified in
the figure.   
The arrows indicate the energy regions where the galactic electrons
dominate and are unaffected by the  Earth's magnetosphere. Around the
geomagnetic cutoff, in the penumbral region, galactic electrons are
mixed with  
re-entrant albedo
electrons that become the dominant component as the energies decrease. }
\label{fig:cutoff}
\end{figure}

The live time was provided by an on-board clock that
timed the periods during which the apparatus was waiting for a
trigger.  The accuracy of the live time
determination was cross-checked by comparing different clocks  
available in flight, which showed a relative difference of less than
0.2\%. The total
live time was about 5$\times 10^{7}$s 
above $\sim$ 20~GV, reducing to about 4$\%$ of this value at 70~MV,
because of the relatively short time spent by the satellite  
at high geomagnetic latitudes. 

Because of the wide geomagnetic region spanned by the satellite over
its orbit, 
the electron energy spectrum was evaluated for various,
sixteen, vertical geomagnetic cutoff intervals, estimated using the
satellite position and the St\"{o}rmer
approximation. Figure~\ref{fig:cutoff} shows the \el~ spectrum
measured in five different geomagnetic regions. Two electron components
can be clearly seen: at energies 
higher than the corresponding geomagnetic cutoff the galactic
component and 
at lower energies the
re-entrant albedo\footnote{Particles produced
in cosmic-ray interactions with the atmosphere with rigidities lower
than the cutoff
that, propagating along Earth's magnetic field line,
re-enter the atmosphere in the
opposite hemisphere but at a similar magnetic latitude.} one, with a
transition region where the 
two components mix. The arrows in Figure~\ref{fig:cutoff} indicate
the energy region (1.3 times above the maximum vertical geomagnetic 
cutoff of each interval, i.e. $1.3 \times 0.93 = 1.209$~GeV for the
black full circle fluxes 
and so on) where the fluxes were assumed to be of galactic origin and 
unaffected by the Earth's magnetosphere. Then, the final electron spectrum
was determined by combining the fluxes of each geomagnetic cutoff interval  
weighted for its fractional live time.

 \begin{figure}[t]
\centering 
\includegraphics[width=14.cm]{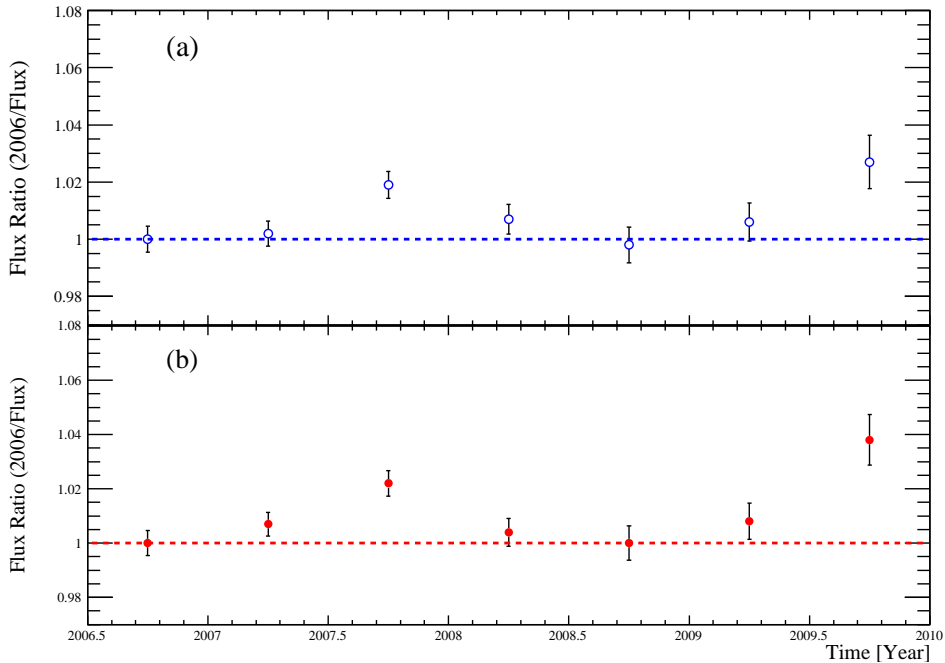}
\caption[]{The high-energy (30-50 GeV) proton flux measured in
July-November 2006 divided by the proton fluxes measured in each time
interval. Proton events were selected with the same requirements of the 
electron analysis  
but the calorimeter selection using 
Criterion 2bis
(a) and Criterion 2 (b).  }
\label{fig:protflux}
\end{figure}

Possible time-dependent variations of the electron fluxes, due to,
e.g., not fully estimated time variations of the tracking selection
efficiencies, were studied as in the proton analysis  
\citep{adr13a}. The high-energy (30-50 GeV) proton flux was measured
for each half year and with the same selections as in this analysis
but the calorimeter selection. The tracking selection efficiencies
were estimated with the same Monte Carlo code used for this
analysis. Then, the resulting fluxes measured in July-November 2006
were divided by the proton fluxes measured in the other time intervals. 
Figure~\ref{fig:protflux} shows this
ratio as a function of time for fluxes obtained with Criterion 2bis
(a) and with Criterion 2 (b). As it can be seen the high-energy proton
flux varies of maximum 2\% over the years with the exception of the
end of 2009 when the flux estimated with Criterion 2 differs of about
4\%. These ratios were used to normalize the electron fluxes measured
using both Criterion 2 and 2 bis in each half-year time interval. \\ 
In conclusion, the final energy spectra were obtained by correcting
the fluxes with these normalization factors and using, as explained in
Section~\ref{s:sel} (see also Section  
\ref{s:sys})\textbf{,} Criterion
2bis up to 500 MeV and the significantly more efficient
Criterion 2 at higher energies.  

\subsection{Systematic uncertainties}
\label{s:sys}
Selection efficiencies were obtained by flight and simulated data
using efficiency samples. The statistical errors resulting from the
finite size of such samples were included in the uncertainties of the
flux measurements and treated as systematic uncertainties. In case of
efficiencies that deviated 
from the fitted values beyond statistical fluctuations
(e.g. see Figure~\ref{fig:Effdedx}), the deviations were observed to
follow a  
Gaussian distribution and the RMS of such distribution was treated
as one standard  
deviation systematic error \citep{dag00}. 

 \begin{figure}[t]
\centering 
\includegraphics[width=16.cm]{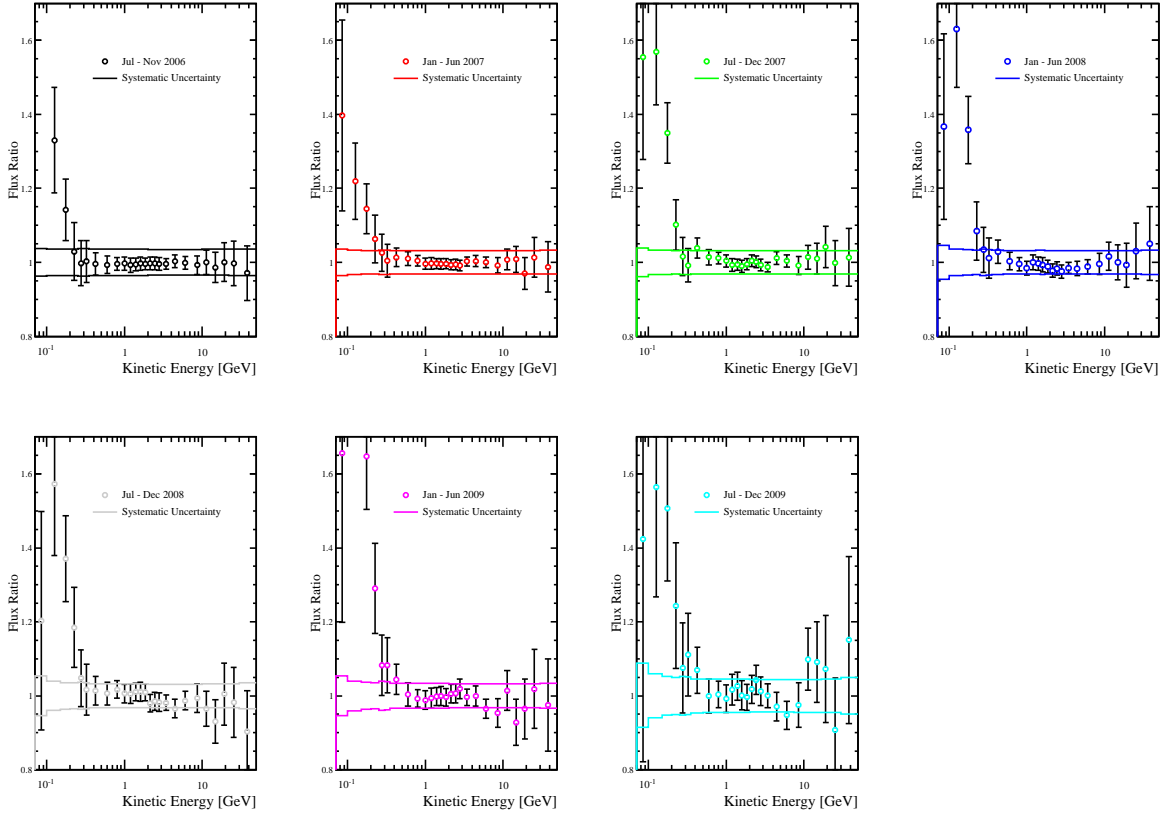}
\caption[]{The electron fluxes measured in each time interval obtained
with Criterion 2 divided by the equivalent ones obtained with
Criterion 2bis.  The solid  lines indicate the
systematic uncertainties associated with these data.  }
\label{fig:334comp434}
\end{figure}

The fluxes were normalized using factors obtained comparing the
high-energy proton flux over time.  
The errors on these factors 
amounted to less than 1\% and were treated as systematic
uncertainties. This normalization accounted for the stability respect
to the second half of 2006 of the fluxes estimated for the following
time periods. A possible systematic uncertainty on the high-energy
proton flux obtained for July-November 2006 and due to the tracking
selection efficiency  was studied as in \citet{adr11b}. An efficiency
sample was obtained both from flight and simulated data selecting
non-interacting minimum 
ionizing particles traversing the calorimeter. This requirement
selected protons with rigidities $\sim 2$ GV and larger. The
resulting simulated and experimental tracking selection efficiency
differed of 1.7\% and 2.3\%  when using Criterion 2 and 2bis,
respectively. Considering that this experimental efficiency sample was
not fully representative of the experimental condition for this
analysis, this difference 
was treated as one standard deviation systematic
error. 

As a check of the consistency of the evaluation of the selection
efficiencies the energy spectrum of each time 
interval obtained with Criterion 2 was compared with the equivalent
one obtained with Criterion 2bis. Figure~\ref{fig:334comp434} shows the
ratios of the two 
sets of fluxes for each time interval. The solid lines indicate the
systematic uncertainties associated with the efficiencies.  
Above 500 MeV, the two sets of fluxes agree perfectly within
the systematic uncertainties showing that systematic errors were
properly assigned to the selection efficiencies. 
Below 500 MeV, the fluxes obtained with Criterion 2 are consistently
higher because of the contamination by spillover protons caused by the
less stringent selection, as discussed in Section~\ref{s:sel}. 

 \begin{figure}[t]
\centering 
\includegraphics[width=16.cm]{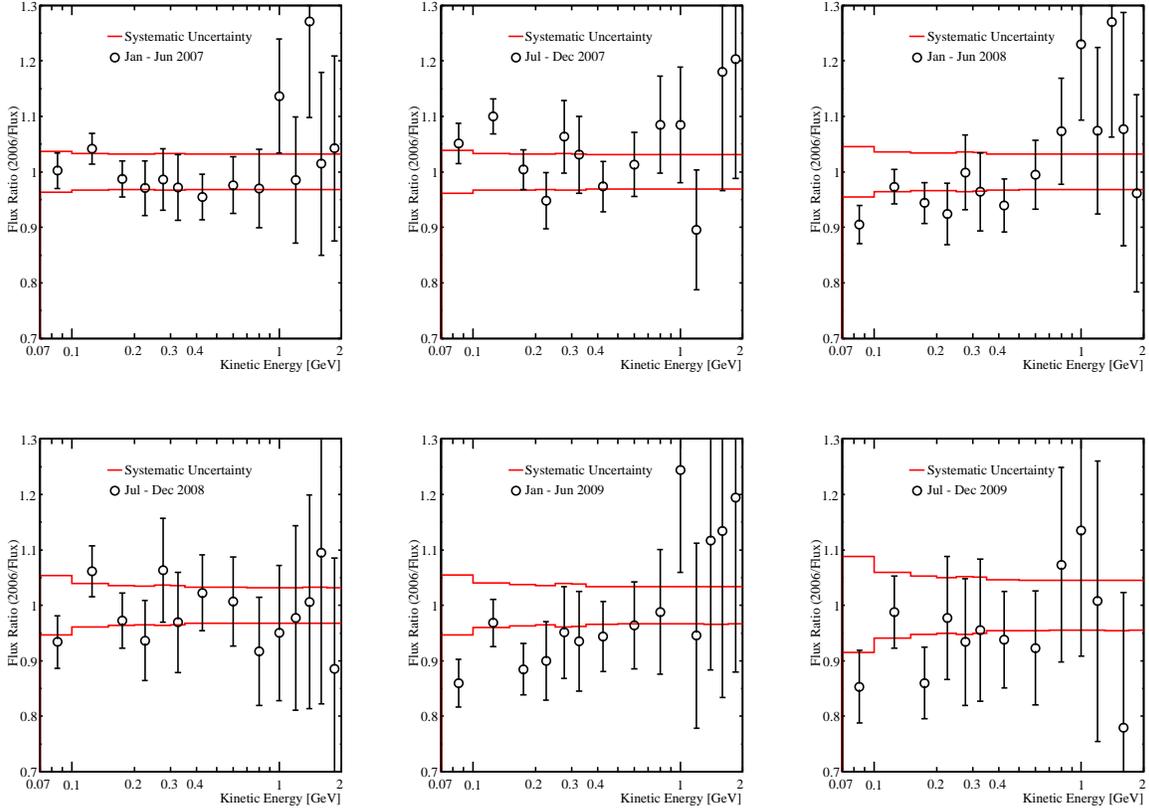}
\caption[]{The re-entrant albedo \el fluxes measured in
July-November 2006 divided by the equivalent
fluxes measured in the other time
intervals.
The solid  lines indicate the
systematic uncertainties associated with these data.  }
\label{fig:albedo}
\end{figure}

An additional check was performed to validate the estimation of the
low energy ($< 1$ GeV) fluxes. The low energy part of the re-entrant
albedo \el spectrum was measured at the lowest geomagnetic latitude
(vertical geomagnetic cutoff greater than 12.1 GV) in each time
interval and it was compared to the same spectrum measured in the
second half 
of 2006. It has been shown \citep{lip02,zuc03} that, because of the
East-West effect, re-entrant albedo \el at low geomagnetic latitudes,
i.e. high geomagnetic cutoffs, are mostly produced by high-energy
($\geq 30$ GeV) protons interacting with the Earth's
atmosphere. Therefore, it can be inferred that the  
re-entrant albedo \el energy spectrum should not show significant
temporal variations due to solar modulation, and hence it can be
used to check the temporal stability of the flux measurements at the
lowest energies.  
Figure~\ref{fig:albedo} shows the re-entrant albedo \el fluxes
measured in
July-November 2006 divided by the equivalent fluxes 
measured in the other time intervals. 
The solid lines indicate the systematic
uncertainties associated with these data. No significant time
variation was found, indicating that the systematic uncertainties
properly accounted for any residual time dependence down to the lowest
measured energies. 

 \begin{figure}[t]
\centering 
\includegraphics[width=15.cm]{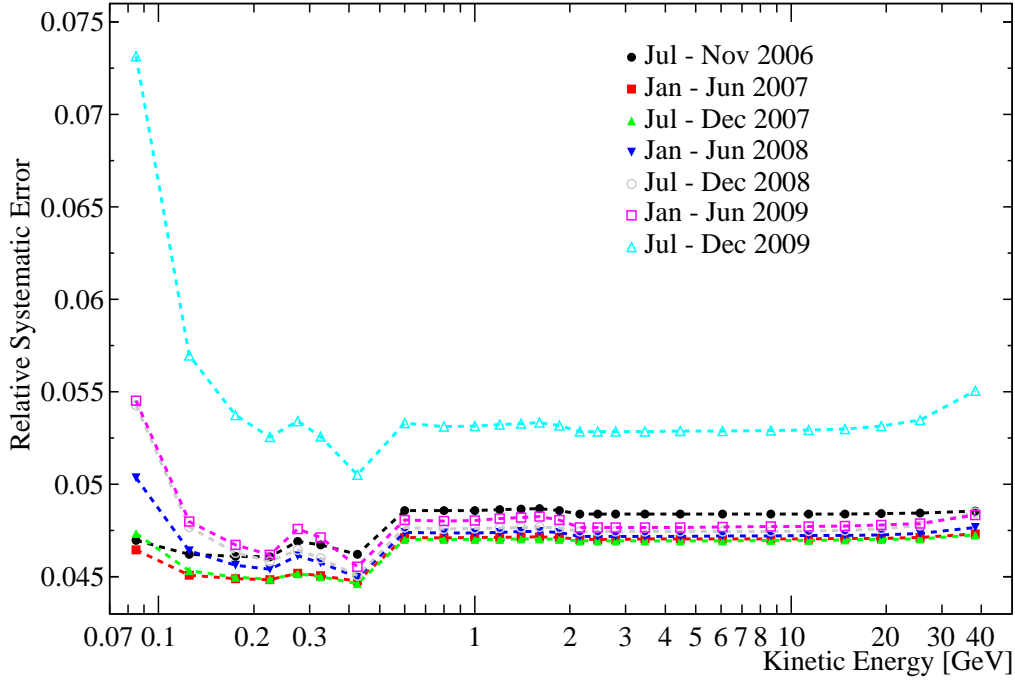}
\caption[]{Relative systematic errors as a function of rigidity for the
  seven time intervals.  }
\label{fig:sys}
\end{figure}

The unfolding procedure was a significant correction for the electron
spectra, therefore the corresponding uncertainties were carefully
studied. It was shown in the proton analysis \citep{adr11b}
that this procedure was able to account for the
intrinsic spatial resolution and the alignment uncertainty of the 
spectrometer silicon sensors. 
The related uncertainties, as well as the 
additional effects due to the significant 
energy losses and reduced statistical significance of the count
distribution, 
were studied by folding and unfolding a known spectral
shape. A large sample of electrons was simulated with
an input spectrum consistent with the reconstructed experimental
spectrum at the top of the payload for the lowest geomagnetic cutoff
rigidity interval.
Then, the rigidities of the simulated events were
reconstructed and one hundred different count distributions were built
as in the analysis. 
The statistics of each count distribution was comparable
with the experimental statistics for a 
geomagnetic cutoff interval. Then, the count
distributions were unfolded and compared with the large simulated
sample by means of pull distributions~\citep{ead71}. These pull distributions
followed the expected standard normal distribution with sigma
consistent with one, hence 
the
statistical errors properly accounted for the fluctuations in the flux
values, and means that fluctuated around zero. 
The relative differences
between the means of the 
expected and reconstructed count distributions  
could be approximated with a Gaussian distribution. Following
\citet{dag00}, the RMS of this distribution, amounting to 4\%, was
treated as one standard 
deviation systematic error due to the unfolding procedure. \\
The unfolding procedure was
also tested comparing the resulting electron energy spectrum with the
one obtained
estimating the electron energy from the total 
energy deposited in the calorimeter (for more information
see~\citet{adr14}). A difference of  2\% at 2 GeV increasing
to 6\% at 10 GeV and then decreasing to less than 1\% above 30 GeV
was found between the two approaches. 
This difference is consistent with the previously estimated unfolding
uncertainty. Hence, even if 
it may also account for additional
uncertainties such 
as those on thickness and density of the materials above the tracking
system, it was not added to the uncertainty of the unfolding
procedure. 

Finally, the full analysis chain was cross-checked with
simulations. Electron 
events were simulated at the top of the payload with isotropic
arrival directions and with energy
spectrum from 40 MeV to 100 GeV consistent with the 
reconstructed experimental
spectrum for the first 
geomagnetic cutoff interval (0-0.055~GV). Then, the events that,
according to 
simulation, triggered
the instrument were processed with the PAMELA data analysis software
and consequently treated  
as in the experimental analysis (rigidity determination, selection
based on Criteria 1-8, efficiency and unfolding corrections, flux
determination). The resulting energy spectrum was compared with the
input one and a good  agreement was
found. The  differences between the input and reconstructed
fluxes at top of the payload were consistent with the 
uncertainties related to the unfolding procedure described in the
previous paragraphs. Therefore, it was concluded that the analysis
procedure did not introduce additional uncertainties.

Figure~\ref{fig:sys} shows the relative errors resulting from the
quadratic sum of the 
systematic uncertainties discussed here. 
Evidently, the uncertainty is higher at
low rigidities, where the Criterion 2bis is used, and it increases
over time, essentially because 
of the decreasing efficiency of the tracking system.

\section{Results}

 \begin{figure}[t]
\centering 
\includegraphics[width=14.cm]{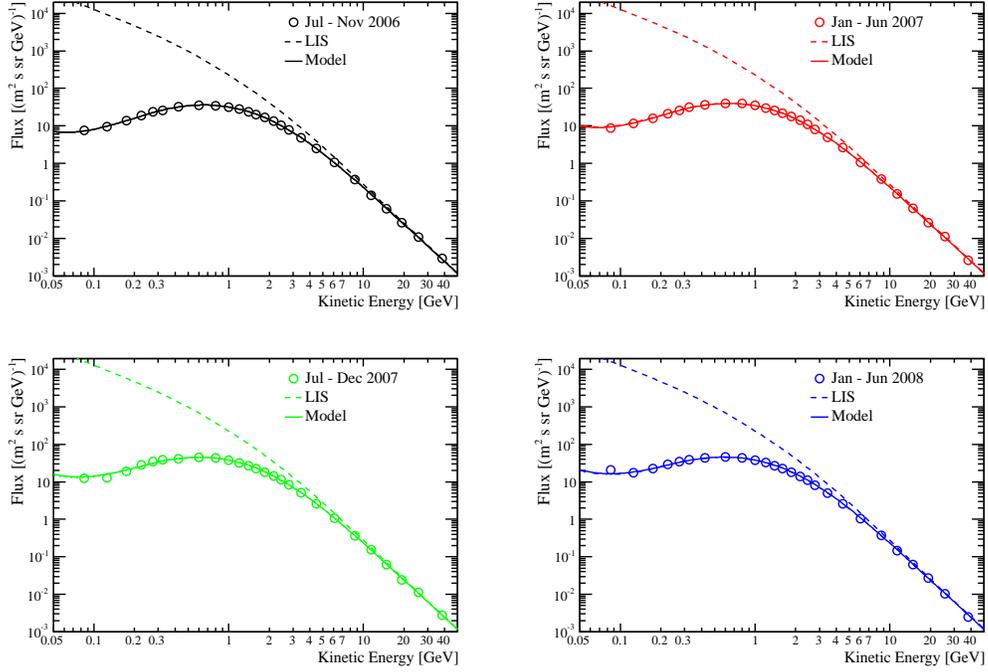}
\caption[]{The measured electron (\el) energy spectrum for the first
half-year periods from the second half of 2006 to the first half of
2008. Time progresses from top to bottom, left to right.
The error bars
are the quadratic sum of the 
statistical and systematic errors. If not visible, they lie
inside the data points.
The computed
spectra (solid lines) and the LIS used for the computation (dashed
line) are also shown.  }
\label{fig:flux1}
\end{figure}

 \begin{figure}[t]
\centering 
\includegraphics[width=14.cm]{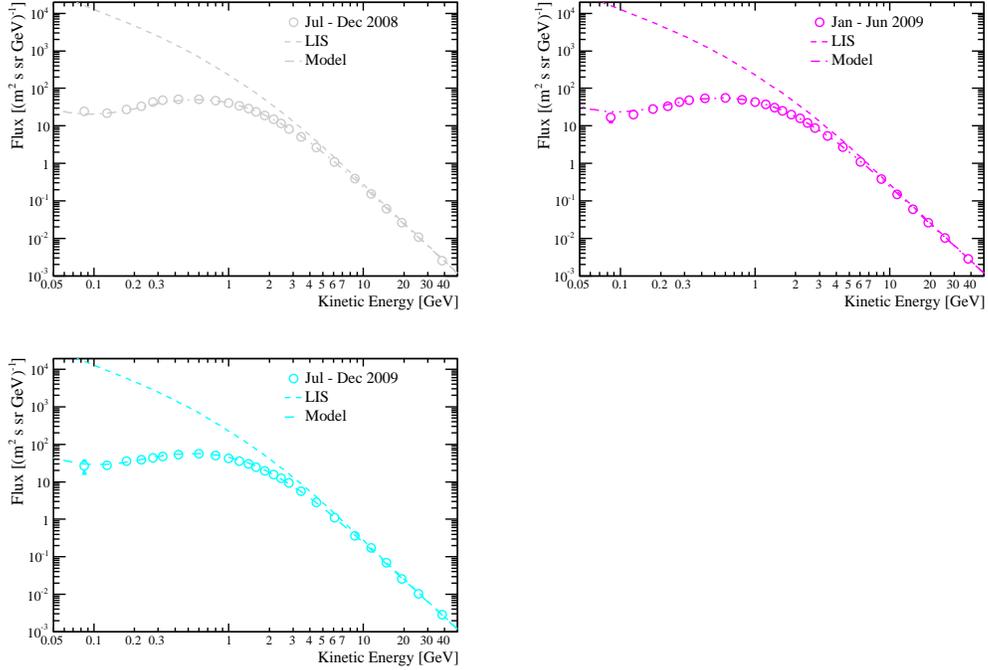}
\caption[]{The measured electron (\el) energy spectrum for the last three
half-year periods from the second half of 2008 to the end of 
2009. Time progresses from top to bottom, left to right. 
The error bars
are the quadratic sum of the 
statistical and systematic errors. If not visible, they lie
inside the data points.
The computed
spectra (solid lines) and the LIS used for the computation (dashed
line) are also shown. }
\label{fig:flux2}
\end{figure}

Figures~\ref{fig:flux1} and \ref{fig:flux2} and Tables~\ref{tab:flux1}
and \ref{tab:flux2} 
show the resulting electron (\el) 
energy spectra for the seven
half-year periods. 
The error bars
are the quadratic sum of the 
statistical and systematic errors.
The electron spectra for each time
interval are overlaid with the corresponding computed spectra (solid
lines) with respect to the local interstellar spectrum (LIS, dashed
lines), which is 
based on Voyager 1 observations \citep{sto13} at low energies. This
LIS was described by \citet{pot13c}; see also the review by
\citet{pot14a}. The full three-dimensional numerical model was
described in detail by \citet{pot14c}. It is based on the numerical
solution of Parker's transport equation \citep{par65}, including all
four major modulation mechanisms: convection, diffusion described by a
full 3D tensor, particle drifts caused by gradients, curvatures and
the current sheet in the heliospheric magnetic field (HMF), and
adiabatic energy changes.

 \begin{figure}[t]
\centering 
\includegraphics[width=14.cm]{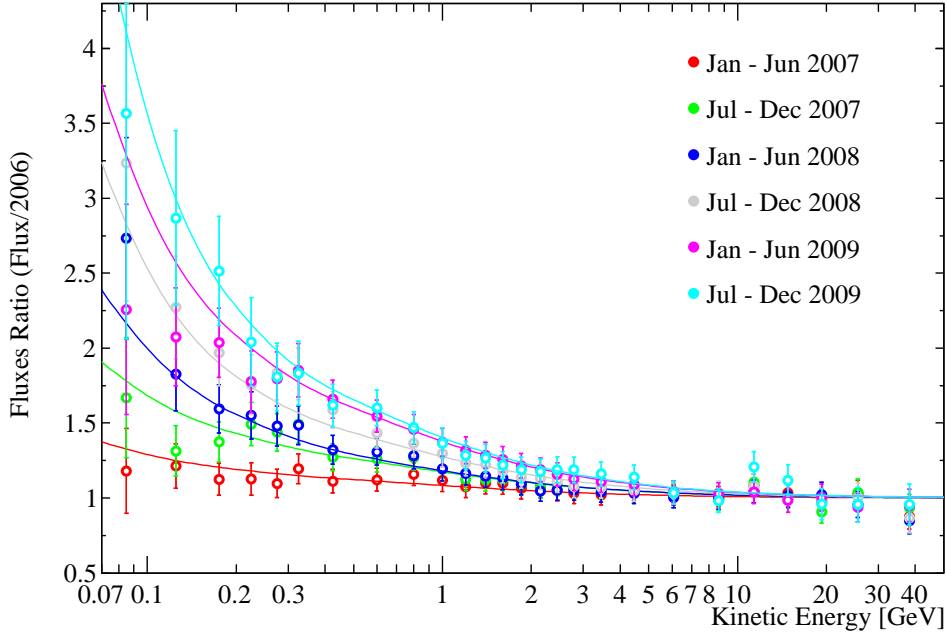}
\caption[]{The ratios as a function of energy 
between the 
measured half-years (\el) fluxes from January 2007 till December 2009 
and the measured fluxes for
the period July-November 2006 
overlaid with the corresponding computed 
spectra (solid lines).
The error bars
are the quadratic sum of the 
statistical and systematic errors. }
\label{fig:ratflux}
\end{figure}

Averaging these fluxes over the whole time period (July
2006-December 2009), the resulting  absolute energy spectrum was 
compared to the previously published results \citep{adr11c}. This new
estimation yields fluxes whose absolute values are approximately 10\%
higher than in the 
previous work. This difference stems from an improved 
treatment both in the data and in the simulation of the time
dependence of the tracking 
system performances and unfolding procedure.   

Figure~\ref{fig:ratflux} shows the ratios as a function of energy 
between the measured half-year
period fluxes from January 2007 until December 2009  
and the fluxes measured in the first period of data taking
(July-November 2006).  
It follows from these ratios that the low-energy electron flux increased
by a factor of about 1.6 from 2006 to 2009 at about 0.5 GeV.  
Protons at corresponding rigidities, on the other hand, increased by
a factor of 
about 2.4 over this period \citep{adr13a,pot14c}, indicating the effect of
particle drifts. Furthermore, 
the comparison between the model simulations and
observations shows that the electron spectrum 
became progressively softer,
more than expected from drift model predictions. This requires larger
diffusion coefficients at lower energies (kinetic energy $<$ 200 MeV)
than anticipated. Details concerning the electron modulation model
with theoretical assumptions and implications will be published in an
accompanying paper \citep{pot15}.  

\section{Conclusions}

We have presented new results on the electron (\el) energy spectrum
between 
70 MeV and 50 GeV obtained
by the PAMELA experiment during the past extraordinary solar minimum
period that ended  
in late 2009 - beginning of 2010. 
By comparing the observations with the model as described in an accompanying
paper \citep{pot15} valuable insight is gained in what caused 
electron modulation over this unusual solar minimum period.

\acknowledgments

We acknowledge support from The Italian Space Agency (ASI), Deutsches Zentrum
f\"{u}r 
Luft- und Raumfahrt (DLR), The Swedish National Space Board, The
Swedish Research 
Council, The Russian Space Agency (Roscosmos) and Russian Science
Foundation. M. Potgieter and E. Vos acknowledge the partial financial
support from the South African Research Foundation (NRF) under the
SA-Italy Bilateral Programme.

\clearpage

\begin{landscape}
\begin{deluxetable}{ccccc}
\tabletypesize{\footnotesize}
\tablecolumns{5}
\tablewidth{0pc}
\tablecaption{Electron flux measured by PAMELA between July 2006 and
June 2008. The first and second errors represent the 
\textbf{1-standard deviation} statistical
and systematic errors, respectively. \label{tab:flux1}    }
\tablehead{
\colhead{Kinetic Energy} &  \multicolumn{4}{c}{Flux} \\
\colhead{(GeV)} &  \multicolumn{4}{c}{(particles/(m$^{2}$ sr
s GeV))} \\ 
\cline{2-5} \\
\colhead{} & \colhead{2006/07-2006/11} &
\colhead{2007/01-2007/06} & \colhead{2007/07-2007/12} &
\colhead{2008/01-2008/06}}
\startdata
0.07 - 0.10  &  (7.48  $\pm$  1.32  $\pm$ 0.35) & (8.83  $\pm$  1.27 $\pm$ 0.41) & (12.48  $\pm$  1.78  $\pm$ 0.59) & (20.43  $\pm$  3.03  $\pm$ 1.03) \\ 
0.10 - 0.15  &  (9.60  $\pm$  0.77  $\pm$ 0.44) & (11.64  $\pm$  0.74 $\pm$ 0.52) & (12.61  $\pm$  0.93  $\pm$ 0.57) & (17.52  $\pm$  1.41  $\pm$ 0.81) \\ 
0.15 - 0.20  &  (13.95  $\pm$  0.75  $\pm$ 0.64) & (15.69  $\pm$  0.70 $\pm$ 0.70) & (19.18  $\pm$  0.92  $\pm$ 0.86) & (22.25  $\pm$  1.21  $\pm$ 1.01) \\ 
0.20 - 0.25  &  (18.74  $\pm$  1.03  $\pm$ 0.86) & (21.11  $\pm$  0.95 $\pm$ 0.95) & (27.99  $\pm$  1.32  $\pm$ 1.26) & (29.07  $\pm$  1.65  $\pm$ 1.32) \\ 
0.25 - 0.30  &  (23.68  $\pm$  1.04  $\pm$ 1.11) & (25.99  $\pm$  0.95 $\pm$ 1.17) & (34.07  $\pm$  1.31  $\pm$ 1.54) & (35.03  $\pm$  1.60  $\pm$ 1.62) \\ 
0.30 - 0.35  &  (26.10  $\pm$  1.05  $\pm$ 1.22) & (31.15  $\pm$  1.00 $\pm$ 1.40) & (38.92  $\pm$  1.34  $\pm$ 1.75) & (38.78  $\pm$  1.61  $\pm$ 1.77) \\ 
0.35 - 0.50  &  (32.56  $\pm$  0.71  $\pm$ 1.50) & (36.23  $\pm$  0.65 $\pm$ 1.62) & (41.44  $\pm$  0.82  $\pm$ 1.85) & (43.01  $\pm$  1.01  $\pm$ 1.93) \\ 
0.50 - 0.70  &  (35.51  $\pm$  0.59  $\pm$ 1.73) & (39.85  $\pm$  0.51 $\pm$ 1.88) & (44.71  $\pm$  0.61  $\pm$ 2.10) & (46.44  $\pm$  0.70  $\pm$ 2.20) \\ 
0.70 - 0.90  &  (34.26  $\pm$  0.42  $\pm$ 1.66) & (39.71  $\pm$  0.38 $\pm$ 1.87) & (43.48  $\pm$  0.44  $\pm$ 2.04) & (43.90  $\pm$  0.50  $\pm$ 2.08) \\ 
0.90 - 1.10  &  (31.26  $\pm$  0.40  $\pm$ 1.52) & (34.90  $\pm$  0.35 $\pm$ 1.64) & (37.25  $\pm$  0.40  $\pm$ 1.75) & (37.37  $\pm$  0.45  $\pm$ 1.77) \\ 
1.10 - 1.30  &  (27.93  $\pm$  0.38  $\pm$ 1.36) & (30.03  $\pm$  0.32 $\pm$ 1.42) & (31.43  $\pm$  0.36  $\pm$ 1.48) & (32.52  $\pm$  0.42  $\pm$ 1.54) \\ 
1.30 - 1.50  &  (23.64  $\pm$  0.28  $\pm$ 1.15) & (26.01  $\pm$  0.24 $\pm$ 1.23) & (26.48  $\pm$  0.27  $\pm$ 1.25) & (27.09  $\pm$  0.31  $\pm$ 1.29) \\ 
1.50 - 1.70  &  (20.00  $\pm$  0.26  $\pm$ 0.97) & (21.96  $\pm$  0.22 $\pm$ 1.04) & (22.47  $\pm$  0.25  $\pm$ 1.06) & (22.64  $\pm$  0.28  $\pm$ 1.07) \\ 
1.70 - 2.00  &  (16.59  $\pm$  0.19  $\pm$ 0.81) & (17.68  $\pm$  0.16 $\pm$ 0.83) & (18.10  $\pm$  0.18  $\pm$ 0.85) & (18.15  $\pm$  0.20  $\pm$ 0.86) \\ 
2.00 - 2.30  &  (13.24  $\pm$  0.16  $\pm$ 0.64) & (14.21  $\pm$  0.13 $\pm$ 0.67) & (14.19  $\pm$  0.15  $\pm$ 0.67) & (13.88  $\pm$  0.16  $\pm$ 0.65) \\ 
2.30 - 2.60  &  (10.41  $\pm$  0.12  $\pm$ 0.50) & (10.94  $\pm$  0.10 $\pm$ 0.51) & (11.31  $\pm$  0.12  $\pm$ 0.53) & (10.97  $\pm$  0.13  $\pm$ 0.52) \\ 
2.60 - 3.00  &  (7.77  $\pm$  0.09  $\pm$ 0.38) & (8.01  $\pm$  0.08 $\pm$ 0.38) & (8.27  $\pm$  0.09  $\pm$ 0.39) & (8.19  $\pm$  0.10  $\pm$ 0.39) \\ 
3.00 - 4.00  &  (4.78  $\pm$  0.05  $\pm$ 0.23) & (4.88  $\pm$  0.04 $\pm$ 0.23) & (5.07  $\pm$  0.05  $\pm$ 0.24) & (4.97  $\pm$  0.05  $\pm$ 0.23) \\ 
4.00 - 5.00  &  (2.51  $\pm$  0.03  $\pm$ 0.12) & (2.64  $\pm$  0.03 $\pm$ 0.12) & (2.60  $\pm$  0.03  $\pm$ 0.12) & (2.58  $\pm$  0.03  $\pm$ 0.12) \\ 
5.00 - 7.50  &  (1.05  $\pm$  0.01  $\pm$ 0.05) & (1.07  $\pm$  0.01 $\pm$ 0.05) & (1.08  $\pm$  0.01  $\pm$ 0.05) & (1.05  $\pm$  0.01  $\pm$ 0.05) \\ 
7.50 - 10.00  &  (3.73  $\pm$  0.07  $\pm$ 0.18) $\times 10^{-1}$ & (3.77  $\pm$  0.06  $\pm$ 0.18) $\times 10^{-1}$ & (3.68  $\pm$  0.06  $\pm$ 0.17) $\times 10^{-1}$ & (3.72  $\pm$  0.07  $\pm$ 0.18) $\times 10^{-1}$ \\ 
10.00 - 13.00  &  (1.41  $\pm$  0.03  $\pm$ 0.07) $\times 10^{-1}$ & (1.53  $\pm$  0.03  $\pm$ 0.07) $\times 10^{-1}$ & (1.56  $\pm$  0.03  $\pm$ 0.07) $\times 10^{-1}$ & (1.48  $\pm$  0.04  $\pm$ 0.07) $\times 10^{-1}$ \\ 
13.00 - 17.00  &  (6.15  $\pm$  0.18  $\pm$ 0.30) $\times 10^{-2}$ & (6.37  $\pm$  0.15  $\pm$ 0.30) $\times 10^{-2}$ & (6.20  $\pm$  0.17  $\pm$ 0.29) $\times 10^{-2}$ & (6.24  $\pm$  0.19  $\pm$ 0.29) $\times 10^{-2}$ \\ 
17.00 - 22.00  &  (2.64  $\pm$  0.10  $\pm$ 0.13) $\times 10^{-2}$ & (2.59  $\pm$  0.08  $\pm$ 0.12) $\times 10^{-2}$ & (2.40  $\pm$  0.08  $\pm$ 0.11) $\times 10^{-2}$ & (2.70  $\pm$  0.10  $\pm$ 0.13) $\times 10^{-2}$ \\ 
22.00 - 30.00  &  (1.08  $\pm$  0.05  $\pm$ 0.05) $\times 10^{-2}$ & (1.11  $\pm$  0.04  $\pm$ 0.05) $\times 10^{-2}$ & (1.12  $\pm$  0.05  $\pm$ 0.05) $\times 10^{-2}$ & (1.03  $\pm$  0.05  $\pm$ 0.05) $\times 10^{-2}$ \\ 
30.00 - 50.00 &  (2.97  $\pm$  0.16  $\pm$ 0.14) $\times 10^{-3}$ & (2.60  $\pm$  0.12  $\pm$ 0.12) $\times 10^{-3}$ & (2.77  $\pm$  0.14  $\pm$ 0.13) $\times 10^{-3}$ & (2.51  $\pm$  0.15  $\pm$ 0.12) $\times 10^{-3}$ \\ 

\enddata
\end{deluxetable}
\end{landscape}

\clearpage

\begin{landscape}
\begin{deluxetable}{cccc}
\tabletypesize{\footnotesize}
\tablecolumns{4}
\tablewidth{0pc}
\tablecaption{Electron flux measured by PAMELA between July 2008 and
December 2009. The first and second errors represent the 
\textbf{1-standard deviation} statistical
and systematic errors, respectively. \label{tab:flux2}    }
\tablehead{
\colhead{Kinetic Energy} &  \multicolumn{3}{c}{Flux} \\
\colhead{(GeV)} &  \multicolumn{3}{c}{(particles/(m$^{2}$ sr
s GeV))} \\ 
\cline{2-4} \\
\colhead{} & 
\colhead{2008/07-2008/12} &
\colhead{2009/01-2009/06} & \colhead{2009/07-2009/12}}
\startdata
0.07 - 0.10  &  (24.19  $\pm$  4.82  $\pm$ 1.31)  &  (16.89  $\pm$  3.99  $\pm$ 0.92) &  (26.68  $\pm$  9.65  $\pm$ 1.95) \\ 
0.10 - 0.15  &  (21.79  $\pm$  2.28  $\pm$ 1.04)  &  (19.91  $\pm$  2.27  $\pm$ 0.96) &  (27.52  $\pm$  4.57  $\pm$ 1.57) \\ 
0.15 - 0.20  &  (27.45  $\pm$  1.94  $\pm$ 1.27)  &  (28.40  $\pm$  2.14  $\pm$ 1.33) &  (35.09  $\pm$  3.95  $\pm$ 1.89) \\ 
0.20 - 0.25  &  (32.92  $\pm$  2.45  $\pm$ 1.51)  &  (33.29  $\pm$  2.61  $\pm$ 1.54) &  (38.23  $\pm$  4.37  $\pm$ 2.01) \\ 
0.25 - 0.30  &  (43.45  $\pm$  2.55  $\pm$ 2.02)  &  (42.50  $\pm$  2.64  $\pm$ 2.02) &  (42.79  $\pm$  4.01  $\pm$ 2.29) \\ 
0.30 - 0.35  &  (47.75  $\pm$  2.54  $\pm$ 2.20)  &  (48.34  $\pm$  2.70  $\pm$ 2.28) &  (47.83  $\pm$  3.97  $\pm$ 2.52) \\ 
0.35 - 0.50  &  (51.70  $\pm$  1.55  $\pm$ 2.33)  &  (54.00  $\pm$  1.70  $\pm$ 2.46) &  (52.67  $\pm$  2.52  $\pm$ 2.66) \\ 
0.50 - 0.70  &  (50.87  $\pm$  0.95  $\pm$ 2.42)  &  (54.92  $\pm$  1.03  $\pm$ 2.64) &  (56.79  $\pm$  1.54  $\pm$ 3.03) \\ 
0.70 - 0.90  &  (46.72  $\pm$  0.68  $\pm$ 2.22)  &  (49.95  $\pm$  0.72  $\pm$ 2.40) &  (50.35  $\pm$  1.05  $\pm$ 2.67) \\ 
0.90 - 1.10  &  (40.51  $\pm$  0.62  $\pm$ 1.93)  &  (42.88  $\pm$  0.66  $\pm$ 2.06) &  (42.72  $\pm$  0.95  $\pm$ 2.27) \\ 
1.10 - 1.30  &  (34.41  $\pm$  0.56  $\pm$ 1.64)  &  (36.86  $\pm$  0.61  $\pm$ 1.77) &  (35.85  $\pm$  0.87  $\pm$ 1.91) \\ 
1.30 - 1.50  &  (28.92  $\pm$  0.42  $\pm$ 1.38)  &  (30.39  $\pm$  0.45  $\pm$ 1.47) &  (29.94  $\pm$  0.65  $\pm$ 1.60) \\ 
1.50 - 1.70  &  (23.62  $\pm$  0.37  $\pm$ 1.13)  &  (25.17  $\pm$  0.40  $\pm$ 1.21) &  (24.38  $\pm$  0.57  $\pm$ 1.30) \\ 
1.70 - 2.00  &  (18.92  $\pm$  0.27  $\pm$ 0.90)  &  (20.15  $\pm$  0.29  $\pm$ 0.97) &  (19.71  $\pm$  0.41  $\pm$ 1.05) \\ 
2.00 - 2.30  &  (14.97  $\pm$  0.22  $\pm$ 0.71)  &  (15.72  $\pm$  0.24  $\pm$ 0.75) &  (15.58  $\pm$  0.34  $\pm$ 0.82) \\ 
2.30 - 2.60  &  (11.57  $\pm$  0.17  $\pm$ 0.55)  &  (12.05  $\pm$  0.19  $\pm$ 0.57) &  (12.37  $\pm$  0.28  $\pm$ 0.65) \\ 
2.60 - 3.00  &  (8.36  $\pm$  0.13  $\pm$ 0.40)  &  (8.75  $\pm$  0.14  $\pm$ 0.42) &  (9.23  $\pm$  0.21  $\pm$ 0.49) \\ 
3.00 - 4.00  &  (5.12  $\pm$  0.07  $\pm$ 0.24)  &  (5.31  $\pm$  0.07  $\pm$ 0.25) &  (5.55  $\pm$  0.11  $\pm$ 0.29) \\ 
4.00 - 5.00  &  (2.63  $\pm$  0.04  $\pm$ 0.12)  &  (2.73  $\pm$  0.05  $\pm$ 0.13) &  (2.85  $\pm$  0.07  $\pm$ 0.15) \\ 
5.00 - 7.50  &  (1.08  $\pm$  0.02  $\pm$ 0.05)  &  (1.10  $\pm$  0.02  $\pm$ 0.05) &  (1.09  $\pm$  0.03  $\pm$ 0.06) \\ 
7.50 - 10.00  &  (3.88  $\pm$  0.09  $\pm$ 0.18) $\times 10^{-1}$ & (3.84  $\pm$  0.10  $\pm$ 0.18) $\times 10^{-1}$ & (3.67  $\pm$  0.14  $\pm$ 0.19) $\times 10^{-1}$  \\ 
10.00 - 13.00  &  (1.53  $\pm$  0.05  $\pm$ 0.07) $\times 10^{-1}$ & (1.47  $\pm$  0.05  $\pm$ 0.07) $\times 10^{-1}$ & (1.71  $\pm$  0.08  $\pm$ 0.09) $\times 10^{-1}$  \\ 
13.00 - 17.00  &  (6.08  $\pm$  0.24  $\pm$ 0.29) $\times 10^{-2}$ & (6.06  $\pm$  0.25  $\pm$ 0.29) $\times 10^{-2}$ & (6.87  $\pm$  0.40  $\pm$ 0.36) $\times 10^{-2}$  \\ 
17.00 - 22.00  &  (2.61  $\pm$  0.13  $\pm$ 0.12) $\times 10^{-2}$ & (2.65  $\pm$  0.14  $\pm$ 0.13) $\times 10^{-2}$ & (2.53  $\pm$  0.20  $\pm$ 0.13) $\times 10^{-2}$  \\ 
22.00 - 30.00  &  (1.08  $\pm$  0.06  $\pm$ 0.05) $\times 10^{-2}$ & (1.01  $\pm$  0.06  $\pm$ 0.05) $\times 10^{-2}$ & (1.03  $\pm$  0.10  $\pm$ 0.06) $\times 10^{-2}$  \\ 
30.00 - 50.00 &  (2.57  $\pm$  0.20  $\pm$ 0.01) $\times 10^{-3}$ & (2.82  $\pm$  0.22  $\pm$ 0.14) $\times 10^{-3}$ & (2.84  $\pm$  0.32  $\pm$ 0.16) $\times 10^{-3}$  \\ 

\enddata
\end{deluxetable}
\end{landscape}

\clearpage

\clearpage

\clearpage

\clearpage

\clearpage

\clearpage

\clearpage

\clearpage

\clearpage

\clearpage

\clearpage

\clearpage

\clearpage

\clearpage

\clearpage

\clearpage

\end{document}